# On the conformance of Android applications with children's data protection regulations and safeguarding guidelines

Ricardo Lopes[1], Vinh Thong Ta[1], and Ioannis Korkontzelos[1]

[1]Department of Computer Science, Edge Hill University, Ormskirk, UK

{lopesr, tav, korkonty}@edgehill.ac.uk

May, 2023

## Abstract

With the rapid development of online technologies and the widespread usage of mobile phones among children, it is crucial to protect their online safety. Some studies reported that online abuse and incidents negatively affect children's mental health and development. In this paper, we examine how Android applications follow the rules related to children's data protection in the EU General Data Protection Regulation (GDPR) and the UK and EU children's online safeguarding guidelines. Our findings show that the number of non-compliant apps is still significant. Even the apps designed for children do not always comply with legislation or guidance. This lack of compliance could contribute to creating a path to causing physical or mental harm to children. We then discuss the relevance of automating the compliance verification and online safety risk assessment, including open questions, challenges, possible approaches, and directions.

## 1 Introduction

The growing use of mobile devices, social media, online gaming, and chat platforms poses a high risk to children's safety. A YouGov study in 2020 found that at least 88% of children in the UK have their own phone by age 12, and at age 6, 85% of children have access to a tablet [1]. Another study [2] in the US examined a set of 336 children and found that the average daily usage among the examined children with their own devices was 115.3 minutes per day. The EU Kids Online research network published a survey [3] in 2020 about tech usage of over 25 thousand children aged 9-16 from 19 European countries and found that up to 45% of children across these countries reported they had been "bothered" or "upset" on the internet.

Unfortunately, many online safety and privacy breach incidents can be read recently in the news. For example, a data privacy breach allegation on how YouTube collects data about children's interaction with the app has been recently filed with the Information Commissioner's Office (ICO). It suggests that Alphabet, the developer of YouTube, collects location and device information as well as data about videos watched by children. Alphabet argues that the app is not designed for children under the age of 13 and that they offer an alternative application for children called YouTube Kids. A similar allegation was made against YouTube in 2019, which resulted in the developer receiving a fine of US$170 million for violating the 1998 Children's Online Privacy Protection Act (COPPA) [4]. In another incident, the findings of the inquest presented by H M Coroner Walker [5] in 2022 and sent to several media and social media organisations, included evidence that harmful social media content contributed to the death of Molly Russell who took her own life [6]. The evidence presented in the inquest confirmed that Molly Russell experienced binge periods of exposure to harmful content, including content that was automatically selected by algorithms designed by social media platforms. Finally, Police Scotland issued a warning about the application Zepeto following an allegation of online grooming. The confirmed investigation involved a child from the Wishaw area allegedly being exploited via the application [7]. After the incident, the

reporting method in the app had been strengthened, introducing anti-grooming features and displaying official safety information.

Given the increasing privacy and safety risk of children online due to the rapid advance of IT technologies, in the future, we anticipate even stricter regulations and enforcement of mandates to protect children than now (examples include the new Online Safety Bill in the UK). Although there are guidelines for businesses on children's online protection (e.g., [8, 9]) and data protection regulations (e.g., [10]), it is unclear how these are adapted/followed by software companies and developers.

Addressing this problem, in this paper, we study how mobile applications comply with data protection regulations and online safeguards with regard to children's safety. We analysed a representative set of 91 Android applications from different countries and of different types against data protection regulations including the General Data Protection Regulation (GDPR [10]), as well as children's online safety guides by the UK Council for Child Internet Safety (UKCCIS) [8, 9], the UK Information Commissioner's Office (ICO) [11], and the ICT Coalition for Children Online [12]. We also examined some US-based apps against the US Children's Online Privacy Protection Act (COPPA) [13], but our main focuses are the EU and UK laws and safeguarding guidelines.

For this purpose, we proposed a systematic analysis method and procedure that contains five steps. The results showed that despite the data protection regulations and safety guides, there are still a significant number of non-compliant Android applications (including apps specifically designed for children). We suggest that automated compliance checks can be a good means to help identify non-compliance at an early development or design stage of the applications, which could be fixed. In this paper, we also discuss the concept of automated compliance verification in this context, its main challenges and potential approaches and future research directions.

Specifically, our contributions are as follows:

1. We review the relevant laws and safeguards related to children's data protection and safety.
2. We define the safety criteria based on the first point to verify the mobile apps against them.
3. We propose a 5-step systematic method to analyse mobile apps regarding children's data protection and safety.
4. We discuss the challenges and findings based on our analysis.
5. We examine the feasibility and challenges of approaches to automating the verification of Android apps.
6. We presented an initial framework for automated conformance verification of Android apps against GDPR legislation and the UKCCIS, ICO and ICT Coalition safeguarding principles.

The rest of the paper is structured as follows: In Section 2, we present a literature review of relevant topics related to mobile app security, privacy, and children's safety. In Section 3, we present the methodology used in this research which involves five steps. The analysis and discussion of our findings can be found in Section 4. In Section 5, we explore the challenges of automating the verification process and present possible approaches to address these challenges. A comprehensive discussion of the findings is presented in Section 6 and finally, in Section 7, we discuss possible future work and present our final conclusions.

## 2 Literature review

Our study focuses entirely on children's data protection and considers EU regulations and EU and UK online safeguarding guidelines in the mobile application context. In the following, we review the most relevant related works.

### 2.1 Security and privacy

According to the study presented by [14], 71% of analysed Android apps that do not have privacy policies ought to have them, and for over 50% of analysed apps that have privacy policies, indications of discrepancies were identified when inspecting their written documentation against their practical operation. It is also suggested that such discrepancies are not committed with malicious intent, but simply due to difficulties experienced by developers in understanding privacy requirements.

In line with the study presented by [14], which reinforces the policies and community standards guidelines published by the ICO [11], our work aims to also verify compliance with age-appropriate guidelines published by UKCCIS [9], specifically in relation to the simplicity of the language used in the documentation aimed at a young audience.

Probably, the most closely related works to ours are the studies by Reyes et al. [15] and Zhao et al. [16], where in the first study, the authors examined thousands of Android apps in the US App Store against the US Children's Online Privacy Protection Act (COPPA), and found that around 57% of the examined apps are potentially violating COPPA. The second study also checked mobile apps (451 apps) against COPPA, but focused on data transmission of children's personal data and found that 67% of the apps showed transmission of identifiers to third-party domains.

The other relevant related works include Kollnig et al. [17], where the authors (manually) analysed a large set of iOS and Android apps for comparing how they ensure (or violate) the privacy of users. Besides privacy properties, security problems were also examined. However, unlike our study, which focuses entirely on children's data protection and safety, the related paper addressed generic privacy and security problems. In our paper, we also analysed the apps against children safety-related safeguards and recommendations, which was not the focus of the study [17].

**Automatic GDPR compliance verification** Following a significant surge in public awareness regarding security and privacy risks as well as the potential worth of personal data, new regulations have been designed and implemented by several governments with the objective of safeguarding the rights of their citizens. However, due to the complexity of certain documents published by app developers, most users do not fully understand the meaning of their agreements. [18] propose an automated GDPR compliance verification tool using an NLP approach with a focus on the detection of regulation violations.

The study published by [18] provides a useful base in the development of our work, however, as it is not directed specifically to children-related legislation and guidelines, our work should propose an improved methodology to automatically verify compliance with children-related legislation, including GDPR and other safeguarding principles.

**Automatic security validation** The introduction of Google Bouncer to Google Play Store in February 2012, which was designed to test submitted applications in a sandbox environment, was intended to provide malware protection by monitoring for potentially harmful behaviours [19], however, developers with malicious intent exploited many vulnerabilities of Google Bouncer due to its inability to provide comprehensive protection [20].

Following a security review convened in 2017, Google Play Protect was launched to replace Google Bouncer. Google Play Protect is not limited to monitoring applications on the server side, as it was designed to be embedded in all Android devices to identify potentially harmful applications on the client side [21].

Our work is different from the approaches taken by Google in their Bouncer software and subsequently, their Play Protect solution, which were both designed to provide malware protection for Android applications. In contrast, our work focuses on the analysis of legislation and guidelines related to children's safeguarding. Unlike Google Bouncer and Google Play Protect, our work is not fully automated. Instead of solely focusing on detecting potentially harmful applications, our work concern the vulnerabilities of children who use these applications and the importance of compliance with regulations and guidance to protect their online safety.

## 2.2 Children and mobile apps

**Child-oriented approach to online safety** Parental mediation is an important protective factor against potentially negative experiences due to exposure to harmful online content. Despite the great number of available parental control applications and the high level of acceptance of those applications among parents, the lack of consideration for the need for privacy is observed in the design concept of that kind of application. Due to a strong personal privacy perception, children and teenagers often show resistance to parental control features, hindering attempts to establish processes designed to increase children's protection online [22].

Due to the focus of our work being directed at ensuring applications follow children-specific legislation and guidelines, we expect to improve parental mediation abilities by reducing the number of non-compliant applications available for children, therefore reducing disruption to the sense of personal privacy and improving the levels of parental control features acceptance among children and teenagers.

**Automating maturity evaluation** The approach to rate the maturity levels of apps is similar to the conventional methods applied to the movie and video game industries, enabling parents to limit access to inappropriate content on their children's devices. However, currently, maturity rating approaches can be expensive or imprecise. Whereas the Apple app store requires maturity ratings to be manually defined by their employees, maturity ratings for apps listed on the Google Play Store are determined by the developers of each application, although Google employees review such ratings in the event of apps being reported for inappropriateness by users. [23] presented a framework based on Machine Learning techniques to evaluate the suitability of individual applications to different age groups, reducing the cost and increasing the accuracy of app maturity evaluation approaches.

Our work proposes a framework that could be used to automatically identify Android software applications that fail to comply with children-related legislation and guidelines, facilitating the process of monitoring applications that could include inappropriate content or functionality. Although not comparable to the complexity of the framework proposed by [23], our work could be used as an additional tool to safeguard children online.

## 2.3 Intelligent safeguard

The widespread use of technology led to a significant transformation in how people interact with each other. While there are numerous benefits to this, there are also negative consequences, such as increased vulnerability to potential dangers online. As a response to this issue, the techno-regulation paradigm is providing innovative safeguarding solutions with the development of tools that can prevent violations of the legislation [24]. Techno-regulation is the term used to describe a set of design choices made to guide and affect human behaviour when interacting with technological resources [25].

Motivated by the necessity of developing a novel solution to protect children by minimising exposure to threats during online activities, [24] proposes a techno-regulatory approach based on machine learning techniques. Their study provides a solution, with an accuracy rate of 88%, for identifying the age of users of mobile applications by analysing touch gestures and distinguishing between underage and adult users.

This study could be a valuable resource for verifying compliance with the first principle in the UKC-CIS guidance [9], which contains recommendations for developers to include age-verification and identity authentication solutions to prevent children from accessing applications that are not age-appropriate. Moreover, utilising this technology could eliminate, or at least minimise, the need to collect personal data from children, which is required in manual age-verification procedures that involve gathering date of birth information.

## 2.4 Our work versus related works

The reviewed literature predominately focuses on the technical aspects of online safety, such as security and privacy, and less on the legal implications of safeguarding children. Seeking to address this gap in the literature, our study focuses entirely on children-related legislation and safeguarding principles. Compared to the studies [15] and [16], our work addresses the GDPR and also consider the EU and UK safeguarding guidelines. Other differences include our proposed 5-step analysis process, different examined app features, and a discussion on automated verification approaches. We hope that our study will raise awareness of the importance of safeguarding children using Android applications and would support decision-making processes associated with the future creation of EU and UK policies and practices in the field.

# 3 Methodology

In this section, we detail the methodology and tools we applied for the compliance analysis of a set of representative Android applications against child-related regulations and safeguards. The regulations and safeguards considered in this study will also be discussed.

## 3.1 Selected Applications

We selected the 91 Android applications based on the following steps. First of all, the list of the Android applications examined in this study includes 23 items selected according to the results attained from internet searches when applying keywords related to apps of high popularity among children and teenagers. These 23 samples were selected from the following websites [26–29]. The remaining 68 apps were selected through the Google Play Store API SerpApi [30] when applying parameters to include only applications suitable for age groups up to 12 years old. This study includes the analysis of applications available for download in the United Kingdom free of charge. Out of the 91 applications selected, 2 were not available for free-of-charge downloads, in which cases, the demo versions were used. The list of the examined applications is distributed across 4 age group ratings and 12 categories, as shown in Table 1 and Table 2, respectively. With the exception of 5 applications, which did not contain an indication of the country of origin registration, the list of countries of origin associated with the applications from the sample examined included 26 countries. The country distribution of the sample is presented in Table 3. The distribution between the 4 age groups across the 12 categories is shown in Figure 1.

Table 1: Age Group Ratings - Applications Sample Distribution [31].

| Age group | Quantity | % | Description |
|---|---|---|---|
| Everyone | 68 | 74.7% | Content is suitable for all ages. May occasionally use mild language and/or minimal amounts of cartoon or fantasy violence. |
| Everyone 10+ | 1 | 1.1% | Content is typically appropriate for ages 10 and up. It could also have more cartoon or fantasy elements, or it could have mild language or minimal suggestive themes. |
| Teen | 14 | 15.4% | Content is typically appropriate for ages 13 and above. It may feature depictions of violence, themes with suggestive content, humour that can be considered vulgar, minimal blood, simulated gambling, and/or sporadic use of strong language. |
| Mature 17+ | 8 | 8.8% | Content is typically appropriate for people who are 17 years or older. It may include scenes of extreme violence, depictions of blood and mutilation, sexual content, and/or the use of offensive language. |

Table 2: Categories - Applications Sample Distribution.

| Category | Total | Everyone | Everyone 10+ | Teen | Mature 17+ |
|---|---|---|---|---|---|
| Education | 25 | 25 | | | |
| Game | 25 | 24 | 1 | | |
| Art & Design | 3 | 3 | | | |
| Health & Fitness | 1 | 1 | | | |
| Music & Audio | 1 | 1 | | | |
| Tools | 1 | 1 | | | |
| Entertainment | 10 | 8 | | 1 | 1 |
| Lifestyle | 2 | 1 | | | 1 |
| Communication | 6 | 3 | | 2 | 1 |
| Social | 15 | 1 | | 9 | 5 |
| Photography | 1 | | | 1 | |
| Video Players & Editors | 1 | 1 | | | |

## 3.2 Considered Legislation and Safeguards

We present the legislation and safeguard principles for children's online safety against which we will analyse the 91 selected applications. In recent years, several pieces of legislation have been enacted to regulate the collection, use, and storage of personal data. One of the most notable examples is the General Data Protection Regulation (GDPR) [10], which came into effect in May 2018 and applies to all EU member states. The GDPR sets out strict rules for the processing of personal data, offering special protection for the handling of children's data. The regulation also grants individuals the right to be forgotten, which allows them to request the deletion of their personal data under certain circumstances. In the UK, the GDPR has been incorporated into the Data Protection Act 2018, which provides additional guidance and requirements for data controllers. The act reinforces provisions related to children's data, such as requiring parental consent for the processing of children's data and setting out rules for the age at which children can provide their own consent [32].

Table 3: Registered Countries - Applications Sample Distribution.

| Country | Everyone | Everyone 10+ | Teen | Mature 17+ |
|---|---|---|---|---|
| <null> | 4 | | | 1 |
| Argentina | 1 | | | |
| Austria | 1 | | | |
| Brazil | 1 | | | |
| British Virgin Islands | 1 | | | |
| Canada | 5 | | | |
| China | 1 | | | |
| Czech Republic | | 1 | | |
| Denmark | 2 | | | |
| France | 4 | | 2 | |
| Germany | 1 | | 1 | |
| Hong Kong | 2 | | | |
| Ireland | 2 | | | 1 |
| Latvia | | | 1 | |
| Netherlands | 1 | | | |
| New Zealand | 1 | | 1 | |
| Pakistan | 2 | | | |
| Romania | 1 | | | |
| Russia | | | 1 | |
| Saint Helena | 1 | | | |
| Singapore | | | 1 | 1 |
| South Korea | 2 | | 1 | |
| Sweden | 2 | | | |
| Turkey | | | | 1 |
| United Arab Emirates | 7 | | | 1 |
| United Kingdom | 5 | | | |
| United States | 21 | | 6 | 3 |

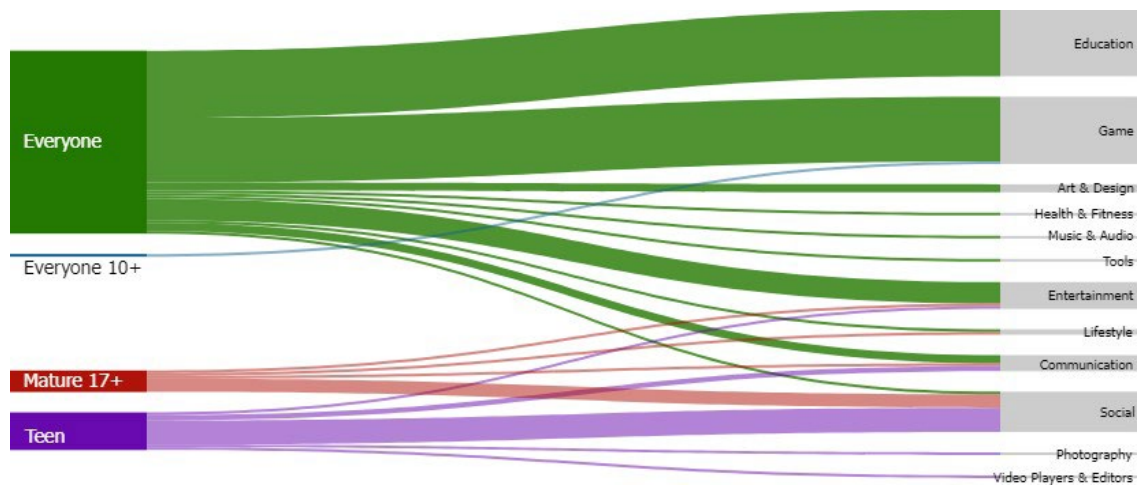

Figure 1: The relationship between age-appropriate content rating and categories of analysed applications. The majority of the 91 apps have been rated as appropriate for all age groups. The number of apps designed for educational and gaming purposes is higher than all other categories, followed by those rated for teenagers, and those rated for mature audiences aged 17 and above, and finally, those rated for ages 10 and above.

In the United States of America, a similar piece of legislation is the Children's Online Privacy Protection Act (COPPA), which was enacted by the US Congress in 1998, and has been operational since 2000. COPPA sets out rules for online service providers who collect personal information from children under the age of 13, including requiring parental consent and providing notice of data collection practices [13].

In addition to legislation, various guidance documents have been published to help organisations safeguard children's online activity. The UK Council for Child Internet Safety (UKCCIS) has produced several resources [9]. The Information Commissioner's Office (ICO) has also published guidance on data protection for schools and childcare providers [11]. The ICT Coalition for Children Online, a European organisation consisting of major technology companies, has published a set of guidelines for the protection of children's privacy online. The guidelines emphasise the importance of providing age-appropriate information to children, obtaining parental consent, and providing clear information about data collection practices [12].

Overall, the combination of legislation and guidance provides a comprehensive framework for protecting children's data and online activity. By following these guidelines, application developers can help ensure that children can safely and responsibly use technology.

### 3.3 Proposed Procedure and Methods

We propose a procedure consisting of five main steps, as depicted in Figure 2, in order to verify the app's compliance with the following aspects of legislation or guidance.

**Step A - Gathering App Details** For the 91 selected applications, we start the process of making a record of their meta information and identification details, such as the logo artwork, their name, product IDs, and the links to their latest versions on Google Store. We then continue with the examination of available documents that lead to the identification of key app characteristics, such as data safety information, ratings and information confirming if the app has been approved by pedagogical professionals.

**Step B - Downloading APK/XAPK and Extracting Content** In preparation for the static analysis process to identify further features in the selected list of applications, the source code must be downloaded from the Google Play Store, and converted to a human-readable format. For this purpose, we download Android Package Kits (APKs) and Android Package Kits with expansion files (XAPKs) that contain the code and resources required for the installation of an Android app.

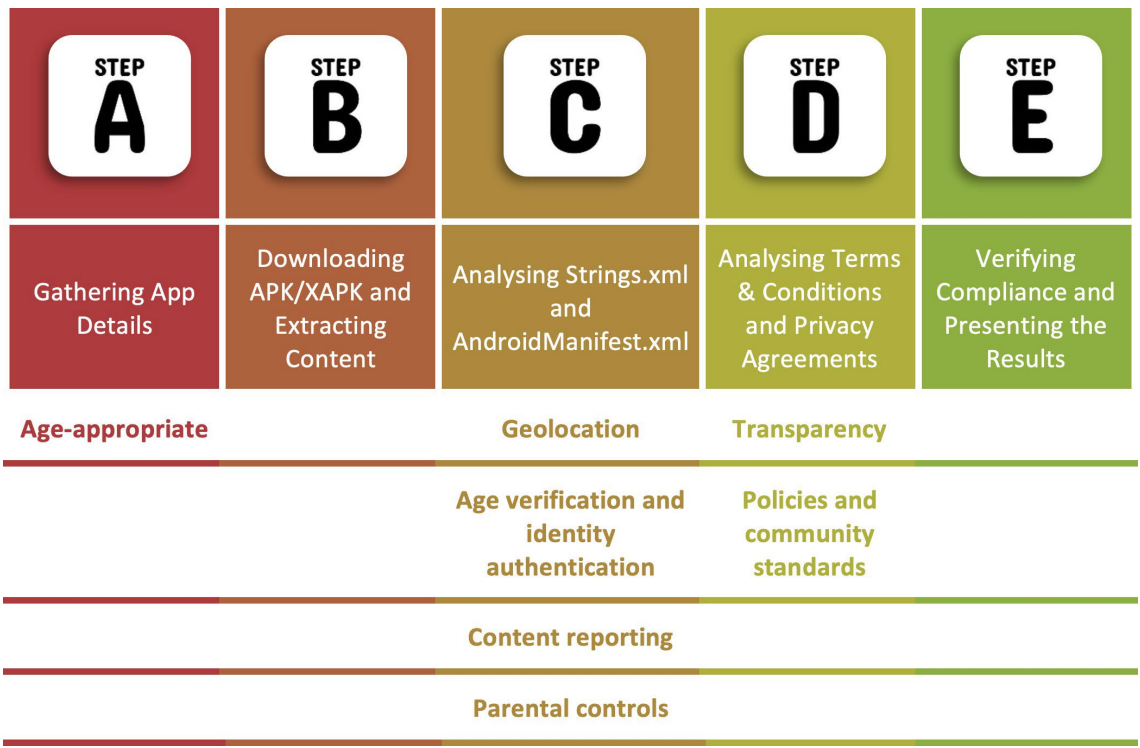

Figure 2: The methodology's five main steps and analysed features. In Step A, while gathering app details, we checked them against the age appropriate concept. In Step C, we analysed the .xml files against the legal requirement on geolocation and three safety principles and guides. Finally, in Step D, we examined the T&Cs and privacy agreements of the apps against the legal requirement of transparency.

**Step C - Analysing Strings.xml and AndroidManifest.xml** An indication of compliance or non-compliance with various guidelines and legal aspects can be attained with the analysis of the extracted and decoded contents of the strings.xml and AndroidManifest.xml files. Android app developers are encouraged to disclose in the AndroidManifest.xml file all app permissions that may be used during user operation. The analysis of the strings.xml files may reveal keywords displayed to the app user, indicating the existence of procedures such as age verification, parental controls, and content reporting.

**Step D - Analysing Terms & Conditions (T&C) and Privacy Agreements** While analysing the stings.xml and AndroidManifest.xml files give us a good overview of the functionalities of the apps, analysing the T & C and privacy agreements of the apps against the functionalities and app operations help us identify any gaps or inconsistencies between these.

**Step E - Verifying Compliance and Presenting the Results** Relying on the extracted app functionalities and the analysis of the T & Cs from Steps C & D, we carry out the compliance verification of the apps against the legislation (GDPR) and the online safety principles presented in Section 3.3.1.

### 3.3.1 The App Features Considered in the Compliance Analysis

In this section, we present and discuss some of the most relevant app functionalities and features regulated by the GDPR (e.g., transparency, consistency with T& Cs, and geolocation data) and addressed by the UK Council for Child Internet Safety (UKCCIS) online safety guides (e. g., age appropriate/children friendly design, parental control, content reporting and age verification).

**Age-appropriate application** The age-appropriate concept requires the perception of the various needs of children at different stages of development and ages, enabling the creation of applications that appropriately serve children that are likely to access the service. Although age ranges do not accurately represent development stages, the Information Commissioner's Office (ICO) has provided guidance to help in the assessment of the skills, capacity and behaviours expected to be displayed by children at each stage of development [11]. The first principle of the Information and Communications Technology (ICT) Coalition states that signatories should clearly indicate services that may include inappropriate content for children, offering access control measures [12]. The UK Council for Child Internet Safety (UKCCIS) provides guidance to ensure that applications that may be used by children are appropriately safeguarded. According to UKCCIS guidance, developers must specify the types of content deemed acceptable, inform such limitations to their users, and clearly define age limits, whilst discouraging access by those considered too young. For applications designed specifically for children under the age of 13, a "walled garden" environment should be considered, in addition to including stricter measures to ensure the privacy of users, including pre- and post-content moderation [9]. Applications awarded the "Teacher Approved" badge have been evaluated by teachers and specialists, under the following criteria: (a) age-appropriate application (interface, content, and adverts), considering the use of language and special effects; (b) attractive visuals; (c) supporting healthy development, including creativity and imagination, learning impact and positive messages, either in the form of audio or written text [33]. Applications with the Teacher Approved badge suggest that the application is age appropriate.

**Geolocation** Recital 38 of the GDPR and the ICO guidance raise concern for children's physical safety when mobile apps collect geolocation data, as in the event of data misuse, children could be vulnerable to risks such as abduction, physical and mental abuse, sexual abuse, and trafficking [11]. App developers can indicate in the AndroidManifest.xml file if their application collects geolocation data.

**Age verification and identity authentication** In addition to considering default safeguards for accounts created by minors, which could contain a higher level of moderation and filtering systems, applications developed primarily for adults could use available age verification and identity authentication techniques, such as checking credit card details, or through an app store account authentication process [9].

**Content reporting** When content or conduct is found to be in breach of a company's Acceptable Use Policy, be it illegal, harmful, offensive, or inappropriate, there should be a clear process for users to report such content or conduct [12]. Additionally, the document presented by UKCCIS includes recommendations for an escalation process in cases of child sexual abuse and illegal sexual contact, alerting the appropriate entities/authorities for further investigation. Information on how to contact any relevant authorities should also be provided [9].

**Parental controls** The guidance provided by the ICO [11] highlights the importance of parental controls, claiming they support parents in the protection of their children; however, it also raises awareness of how parental controls may impact children's rights to privacy, possibly affecting their development. To minimise the potential negative impact on children, the ICO recommends that developers ensure that children from different age groups are clearly informed when being monitored or tracked. In addition, while the ICT Coalition proposes the provision of assistance to parents wishing to prevent their children from accessing potentially harmful content [12], UKCCIS guidance highlights the importance of ensuring any parental control tools are easy to use [9]. Such tools must be appropriately tailored to the products or services offered and should not be expected to substitute parent engagement when limiting online access [12].

**Transparency** The concept of transparency can be enforced by Article 12 of the General Data Protection Regulation (GDPR), which states that information associated with the processing of personal data must be provided in a transparent, concise, intelligible, and easily accessible form, using easy-to-understand language, especially when addressing children [10]. In addition to highlighting the importance of the legislation included in the GDPR, guidance provided by the ICO, ICT Coalition and UKCCIS includes implementation details. According to UKCCIS, the importance of being clear on the minimum age limits could help create a safer environment for users intending to share content [9]. The ICT Coalition encourages developers to clarify details of the consequences to be sustained by users that fail to adhere to what is deemed as acceptable behaviour when utilising the application [12].

**Policies and community standards** The guidance provided by the ICO emphasizes the importance for app developers to adhere to their own policies, published terms and community standards [11]. Such a recommendation is supported by article 5(1) of the GDPR, which states a breach of regulation if a service is operated differently from what is published on the developer's terms and conditions, especially when personal data is collected from children [10].

## 3.4 Technical Discussion

In this section, we discuss how the steps A-E of our procedure were carried out, detailing the used tools and methods.

### 3.4.1 Step A: Gathering App Details

The process to gather software application details included manual searches for online articles classifying Android applications of high popularity among children and teenagers. The search was focused on websites specifically designed to influence the opinion of children and teenagers, in hope to identify sources used by children and teenagers seeking specialised software applications.

Additionally, the Google Search API SerpApi was used to collect details of software applications designated for children. We combined the results collected when applying the parameters listed in Table 4 with the results from the manual searches and store them in a single JavaScript Object Notation (JSON) file. Each entry in the JSON file contains the following information related to an application: (a) *title* (the official application title); (b) *link* (the direct web link to the application listing in the Google Play Store); (c) *product_id* (the unique identification code); (d) *serpapi_link* (the direct web link to a document, stored within SerpApi servers, containing additional information about the application, such as the developer's contact details and similar products); (e) *thumbnail* (the direct web link to the official image icon of the application); (f) *rating* (the app's grade rating (0-5), as shown in the Google Play Store at the time of the search, although, this information is omitted for the applications with no user feedback and rating data).

Table 4: Parameters used in the search conducted via SerpApi.

| Parameter | Value | Description |
| --- | --- | --- |
| engine | google_play | Search conducted in the Google Play Store. |
| store | apps | The type of items searched, excluding games, movies and books. |
| apps_category | FAMILY | This subcategory defines apps designed for age ranges up to 12. |
| hl | en | Definition of English as the search language. |
| gl | uk | Definition of United Kingdom as the country. |
| api_key | N/A | The API private access key. |

**Age appropriateness** In order to verify the age appropriateness of applications we access the direct web link to the application listing in the Google Play Store. For the verification, we search for the keywords "Teacher Approved" in the code. The keyword "Teacher Approved" is included in the listings of applications that comply with rigorous guidelines imposed by Google to promote their suitability for children.

### 3.4.2 Step B: Downloading APK/XAPK and Extracting Content

In the next step, we extract the content and information about the apps by downloading their APK or XAPK files. Android Package Kits (APK) and Android Package Kits with expansion files (XAPK) contain the code and resources required for the installation of an Android app via the Google Play Store [34]. The process of downloading such files can be achieved through one of the various freely available websites offering such services, including Evozi [35], APK Pure [36], APK DL [37], and APK Mirror [38]. Following the download of the APK files, a reverse engineering tool, called the APK Tool [39] was used to decode Android application resources with the intention of analysing key features offered by the selected applications. Subsequently, two files were extracted from the source code: Strings.xml and AndroidManifest.xml.

### 3.4.3 Step C: Analysing Strings.xml and AndroidManifest.xml

To identify and extract the functionalities and features of the apps as the next step we carry out a static analysis (searching for keywords) of the strings.xml and AndroidManifest.xml files obtained from the previous step.

In the structure of an Android Studio project, the Strings.xml file is usually located in the folder "values"/"res" and it contains text attributes and values for different GUI elements used in the apps, such as text views, buttons, text fields, checkboxes and radio buttons. These can be referenced at different locations of the code, which makes it more convenient to avoid repetitive definitions of the same text in large applications. Although Strings.xml may not include all the necessary information, it generally contains a substantial amount of information about the functionalities and features of an app. The challenge then lies in identifying these features using data analytics approaches, as different apps may refer to the same functionalities with different names.

The AndroidManifest.xml file is required for all Android applications and can be found in the root directory of an Android Studio project. It contains information about the package, including its various components such as activities, services, broadcast services, and content providers. The file also defines the permissions necessary for the application to access protected parts of the system and declares the Android API that the application will use. Additionally, it lists the instrumentation classes that provide profiling and other information.

In the following, we will discuss how the features and functionalities presented in Section 3.3.1 were identified and examined in the Strings.xml and AndroidManifest.xml files. These were done in an automated way using our Python scripts [1].

**Geolocation** To check if an app requires permission to access geolocation data the followings were done: (a) First, we opened the AndroidManifest.xml file and extract its contents; then (b) from the contents extracted from the AndroidManifest.xml file, the special characters (such as <"-\_:=./>[]) were removed; then (c) we searched the file for the strings that contain the "location" keyword such as "access fine location" and "access coarse location"; (d) to identify additional relevant keywords, we used WordNet, a lexical database for English by Princeton University [40], to generate synonyms of the basic strings to increase the chance of finding matching information; finally (d) based on the set of keywords and strings, we carried out the search for these in the AndroidManifest.xml file.

**Age verification and identity authentication** The process to check if an app includes at least one form of the age verification process is similar to the previous case, but the Strings.xml file was examined instead of the AndroidManifest.xml file. In addition, we searched for a set of keywords and strings related to "age verification". Again, similar to the first case, WordNet was used to attain additional sets of keywords/stings by identifying synonyms based on part-of-speech equivalency.

**Content reporting** In the case of content reporting, similar to the case of age verification, we searched for keywords and strings in the Strings.xml file, but used the predetermined keywords in the Table 5. We set up this table manually with the keywords and strings that refer to the most frequently found inappropriate content.

**Parental controls** For the parental control features, we followed the same process as the previous cases to see if Strings.xml files contain the keywords/strings containing "parental consent" and "parentalconsent" (and any of their synonyms based on WordNet).

### 3.4.4 Step D: Analysing Terms & Conditions and Privacy Agreements

In the fourth step of our methodology, we analysed the terms & conditions and privacy agreements of the 91 Android applications from our list. In Step D, as depicted in Figure 2, we aimed to verify the compliance of each app against the legislation and guidance related to children, with a focus on key issues such as transparency and policies and community standards. Due to the lack of a standardised format and the complexity of these documents, this step was completed by manually checking their contents.

---

[1]Our Python scripts can be found here: https://github.com/rics23/ChildDataVerif

Table 5: Content report - the predetermined array of keywords/phrases we used.

| block and report | block or report | report abuse | report as inappropriate |
|---|---|---|---|
| report bullying | report comment | report content | report explicit image |
| report extremism | report hate speech | report imminent danger | report inappropriate |
| report nsfw | report nudity | report or block | report pornograph |
| report sexually explicit | report this contact | report this group | report this member |
| report this photo | report this post | report this user | report this video |
| report user | reportchatchild | reportchatpornography | reportchatviolence |
| reporting harassment | reporting hateful | reporting nudity | reporting self harm |
| reporting violence | thanks for reporting | the post you reported has been removed | |

### 3.4.5 Step E: Verifying Compliance and Presenting the Results

The last step in our methodology involves cross-referencing the observed functionalities of the app with any legal or regulatory constraints. This verification has the potential to provide a comprehensive list of compliance and non-compliance features of the 91 analysed apps. Once the verification process is complete, the results of the analysis were summarised in a clear and concise manner to present the most relevant information attained.

## 4 Findings

We discuss the results of the analysis, which we divided into two subsections, namely, compliance with safeguard principles and compliance with regulation/legislation. In each case, we follow the app features presented in Section 3.3.1. In the following, out of the 91 applications we examined 69 applications that were rated as suitable for children under 13, and 22 apps were designed for older audiences. The complete Excel table containing our analysis results can be found on the GitHub page of the project [2].

### 4.1 Compliance with the Safeguard Principles

In the following, we discuss the compliance of the examined apps with the safeguard principles by UKCCIS [9].

**Age-appropriate guidance** Based on our methodology above, we found that 4 out of the 69 apps designed for children suggested a lack of compliance with the age-appropriate guidance provided by the ICO [11], ICT Coalition [12] and UKCCIS [9]. Furthermore, we found that 1 out of those 4 applications even stated that (with permission) the collected personal data can also be shared with third parties. However, sharing children's personal data with third parties, without a convincing reason to do so in the best interests of the child can be seen as noncompliance with the ICO guidance [11].

**Geolocation** According to the ICO guidance [11], geolocation should be switched off by default, unless the developer is able to demonstrate a compelling reason for the operation of the app, and geolocation data collection should be switched off at the end of each session. Upon the employment of our methodology, evidence suggesting that geolocation permission is requested by default was observed on 6 out of the 69 apps rated for children under 10, illustrated in Figure 3. Out of these 6 apps collecting geolocation information, 3 were listed as communication apps, which could be a compelling reason to collect fine and coarse location data, however, the remaining three apps are categorised as entertainment, health & fitness, and tools, indicating a possible breach of the ICO guidance.

**Harmful content reporting** We found that the content reporting features were missing in 65 out of the 69 apps designed for children under 13, and only in 3 out of the 22 apps designed for older audiences. While at first sight, this may look surprising, the contents of the apps designed for children probably already followed safeguards and guides, and therefore the developers may think that harmful content reporting is unnecessary, as this feature is only recommended safeguards (by the ICT Coalition and UKCCIS) and not mandatory requirements. Among the 22 apps designed for older audiences, we identified that the 3 apps with no content reporting features are classed as social applications, which generally allow users to broadcast unsupervised content, potentially enabling the publishing of harmful content.

---

[2]The complete Excel table containing our analysis results: https://github.com/rics23/ChildDataVerif

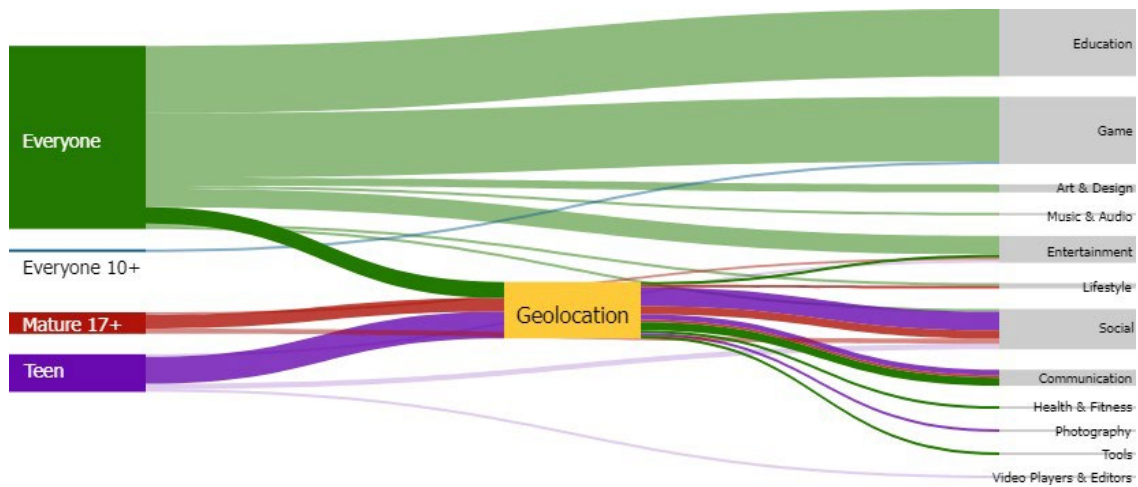

Figure 3: The distribution between age groups and categories, highlighting geolocation access apps.

**Parental control** According to the guidance provided by the ICT Coalition, developers of applications designed for older users should provide parental control features to prevent exposure to inappropriate content and contact. Based on our methodology, less than 14harmful content included indications of parental controls as an incorporated feature of the apps. Indications of parental control features were found in 38 out of the 69 apps designed for children under 13, and in only 3 out of the 22 apps designed for users older than 13. The lack of parental controls in the apps designed for children does not necessarily suggest regulation or legislation breaches, as despite this, the potential for inappropriate content within apps designed for a younger audience can be minimal. On the contrary, the lack of this feature in apps designed for older users raises concerns as often there are not enough measures to prevent minors from using them.

**Analysing Terms & Conditions and Privacy Agreements** Overall, our analysis of the privacy agreements and T & C revealed that a majority of the documents did not meet an acceptable level of transparency. Moreover, it suggests that developers may struggle to enforce their own policies and community standards practically. It is worth noting that the agreements we analysed may have been updated since the time of our analysis.

We analysed a range of T & Cs and privacy agreements, where some were informative and transparent, and the others were either vague or highly technical. For example, as shown in Table 6, the app with the ID A43 used colourful and child-friendly language to explain what information is collected and how it is used, including the sharing of data with third parties. The agreement also mentioned local legislation and seemed to adhere to policies and community standards. A68 was also transparent in terms of what and how personal data are collected but was written in a technical language which is not entirely child-friendly. However, it did include a section for children explaining how their data is used and appropriate parental consent procedures.

## 4.2 Compliance with Legislation and Regulations

In this section, we discuss the compliance of the examined apps with the GDPR.

Table 6: Apps collecting personal data without offering a procedure to erase personal data. (We took out the names of the apps as they are irrelevant to the results and replaced them with unique IDs from A01-A91).

| App IDs | Age rating | Category | Data collected? | Right to erasure? | Types of data collected |
|---|---|---|---|---|---|
| A05 | Everyone | Education | ✓ | ✗ | Location, app activity, app info and performance, device and other IDs. |
| A23 | Everyone | Education | ✓ | ✗ | **Name, email address**, **user IDs, phone number**, **emails**, app activity, app info and performance, device or other IDs |
| A24 | Everyone | Education | ✓ | ✗ | **Name, email address**, **user IDs, phone number**, **emails**, app activity, app info and performance, device or other IDs. |
| A25 | Everyone | Education | ✓ | ✗ | App activity |
| A43 | Everyone | Game | ✓ | ✗ | App activity, app info and performance. |
| A44 | Everyone | Game | ✓ | ✗ | Location, app activity, device or other IDs |
| A46 | Everyone | Game | ✓ | ✗ | Location, app activity, device or other IDs. |
| A52 | Everyone | Game | ✓ | ✗ | App info, performance |
| A70 | Teen | Communication | ✓ | ✗ | **Email address and phone number**, **messages**, **photos and videos**, **contacts**, app activity, app info and performance, device or other IDs. |
| A89 | Mature 17+ | Social | ✓ | ✗ | Location, **email address financial info, messages**, **photos and videos**, app activity, app info and performance, device & other IDs |

**Right to be forgotten** According to Article 17 of the GDPR [10], organisations collecting personal data should offer a procedure guaranteeing the clients their right to be forgotten, however, according to statements issued by the developers of 10 apps from the analysed set (see Table 6), permission to collect personal data is required without the assurance that the deletion of such data is presented. In Table 6 we took out the name of the apps and replace them with unique IDs between A01-A91 as we think that the names of the apps are irrelevant to the analysis results. Instead, we included the categories and age ratings of the apps.

**Age-appropriate guidance** Sharing children's personal data with third parties, without a convincing reason can also be seen as noncompliance with Article 8 of the GDPR. As mentioned in the previous section, we found that 1 out of those 4 applications (that suggested a lack of compliance with the age-appropriate guidance) allows the collected personal data to be shared with third parties.

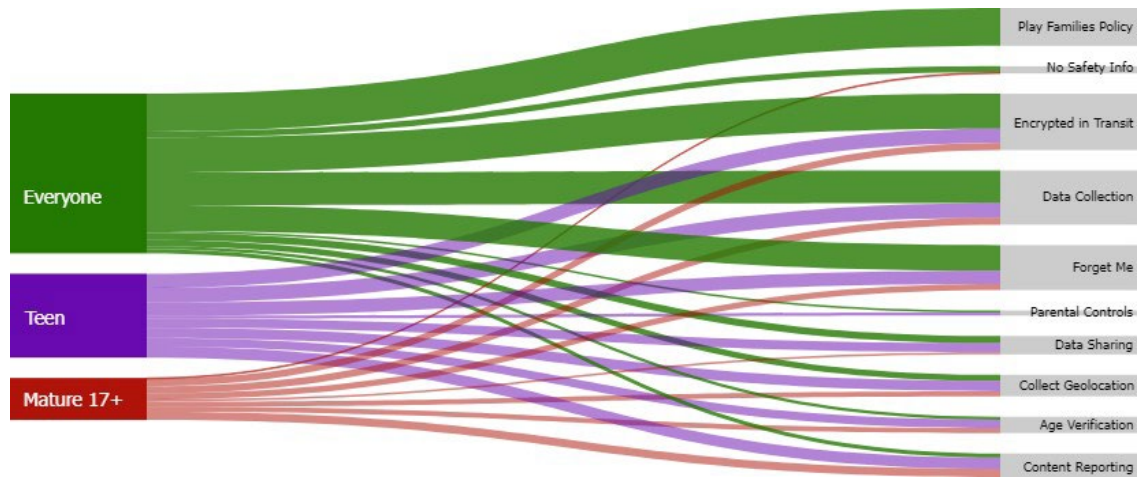

Figure 4: Distribution between age group ratings and identified features.

**Geolocation** The Recital 38 of the GDPR discusses the importance of protecting children's personal data. While children are using mobile devices the collected geolocation data becomes their personal data. The Recital 38 of the GDPR discusses the importance of protecting children's personal data. While children are using mobile devices the collected geolocation data becomes their personal data. As mentioned in the previous subsection, we found that 6 out of the 69 apps rated for children under 10 requested permission for the collection of geolocation data.

**Analysing Terms & Conditions and Privacy Agreements** While analysing the T & Cs and the privacy agreements of the 91 apps, we assigned the apps with unique IDs between A01-A91. We identified that in the case of some applications such as A20 (a US-based education app for all age groups in the US), it stated that guest users are not required to provide personal information and that network data is anonymised, and it meets the transparency requirement of both GDPR and COPPA, given that this is a US-based app. In contrast, the privacy agreements of some apps lack details such as the app A31 (a French-based education app for all age groups) only stated that the developer does not collect user data, while the app A09 (an entertainment app for all age groups) lacked details related to personal data storage and processing. In addition, the app A40 (a Pakistan-based entertainment app for all age groups) included guidance stating that any information submitted by children would be refused or safely deleted, but does not fully meet the transparency requirement of the GDPR, and did not include a professional email address.

### 4.3 Remark

Out of the 7 app features and functionalities in Section 3.3.1 that we examined we found non-compliant apps in the cases of 5 features, namely, age appropriate geolocation, harmful content reporting, transparency (in the analysis of T & Cs and privacy agreements) and parental controls. We also analysed the 91 applications against the features of Age Verification and Identity Authentication, and Policies & Community Standards, where we did not identify any non-compliance.

## 5 Automating the verification

Based on our analysis of a representative sample of Android applications, it can be seen that despite the attempts and willingness to adhere to the data protection regulations and safeguard principles, there are still many non-compliant applications. Therefore, it is desirable to develop an approach or software tool that can be used to identify non-compliance in the applications in an automated way. Addressing the future regulations (e.g., the prospect of the UK Online Safety Bill), automation would help app developers and designers be aware of the non-compliance problems at an early stage to fix them, and also non-compliance can be used in the process to design warnings for parents and children. This ultimately would reduce the rate of non-compliant applications and improve children's safety online.

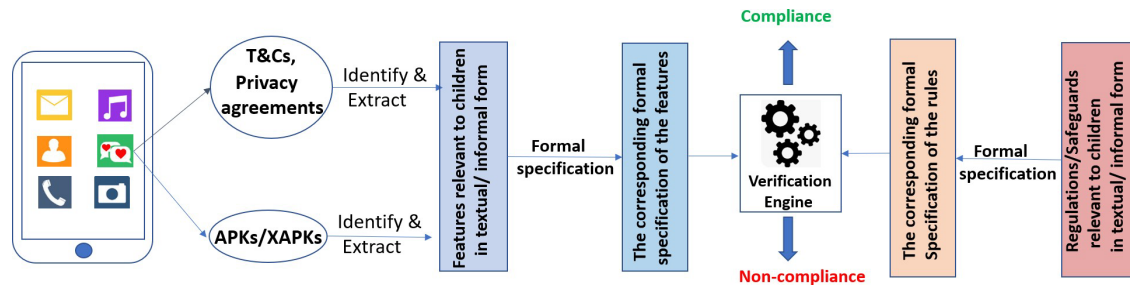

Figure 5: A concept of automated compliance verification. The verification engine can be based on automated formal proofs (model-checking and theorem proving) or machine learning classification.

A concept of automated compliance verification is depicted in Figure 5. Following the methodology presented in Figure 2, the T & Cs and privacy agreements of the apps are analysed and the features/functionalities relevant to children's online safety discussed in Section 3.3.1 are identified and extracted. These are then saved in a textual (informal) form, and checked against the relevant regulations and safeguard principles. As by default, both the extracted app features and the relevant rules/safeguards are in textual format, if the automatic verification is done directly on them, then the result can be inaccurate. An approach to making the verification more accurate is to provide formal specifications of the features and the rules/safeguards. In this case, compliance can also be proved using sound mathematical approaches (with model-checking or theorem-prover tools). Alternatively, based on the app features machine learning methods specifically designed for this context, can be used to classify if an app violates a rule/guideline. However, this latter approach may produce high rates of false negatives/positives. In any case, (fully) automatic compliance verification of applications is a very challenging task. In the following, we will discuss the main challenges.

## 5.1 Challenges

Based on the steps/tasks in Figure 5, the challenges can be grouped into five categories: (i) automatic identification and extraction of relevant features from the T & Cs and privacy agreements, (ii) automatic identification and extraction of relevant features from APKs/XAPKs, (iii) formal specification of the features, (iv) formal specification of the rules/safeguards, and (v) the automated verification method/engine.

**Challenge I - Automatic identification and extraction of features from the T & Cs and privacy agreements** Terms & conditions and privacy agreements are lengthy text documents that can contain an enormous variety of phrases and combinations of words and sentences in English or other languages. Automatically analysing these text documents to detect parts of the text that are relevant to children's data and online safeguards is a very challenging and complex task. To make the identification accurate, the semantics or meaning of the text phrases and combinations of words need to be modelled. There are related works in the literature based on natural language processing and machine learning to analyse privacy policies and contractual documentations (e.g., [41–43]), but their proposed approaches only analyse a limited set of phrases and are not related to the context of children safety.

**Challenge II - Automatic identification and extraction of relevant features from APKs/XAPKs** As mentioned in Section 3.4, the analysis of the strings.xml and AndroidManisfest.xml files is based on keywords and phrases identification, hence, we face a similar challenge as in T& Cs. However, beyond these, the analysis of the app source codes would be necessary to provide a more comprehensive picture of the functionality of an application, and a more accurate verification. In general, we distinguish between static and dynamic code analysis. Static code analysis focuses on analysing the software codes without the need to execute them. In our context, it would mean that the syntax and semantics of the codes are analysed to identify relevant features and functionalities for children's online safety. While static analysis has been an intensively researched area [44–46], applying it in the area of children's online safety is less studied and comes with new challenges. Finally, another challenge to face is that many applications have obfuscated codes (to protect intellectual property), which is a great obstacle to code analysis.

Dynamic code analysis is based on evaluating the actual behaviour of mobile applications in order to identify any discrepancies between their expected functionality and their actual operation. The set of features extracted during dynamic analysis complements the ones in static analysis. There are several automated testing tools for mobile apps such as Appium [47], however, developing an approach to automatically identify and extract relevant features to children's safety is a new challenge.

**Challenge III - Formal specification of the app features.** Once we have the extracted app features and functionalities, for automated verification, we need to model these features in machine-understandable format. To address this, we need an appropriate formal language with accurate definitions of the syntax and semantics of the app features/functionalities. To the best of our knowledge, no formal language has been proposed for this purpose.

**Challenge IV - Formal specification of the GDPR rules and safeguards.** Similar to the previous case, formal language is required for this purpose. In [48], we proposed a formal language for specifying a set of GDPR rules and system architectures to verify their conformance. However, it focuses mainly on generic GDPR rules rather than rules related to children, and it does not consider mobile app features.

**Challenge V - The automated verification method/engine.** After defining a set of formal specifications of app features and regulations we now need an algorithm to verify them "against each other". The main challenges, in this case, are to define appropriate verification goals, specify the connection between the two semantics, and automate the mathematical proofs of compliance/non-compliance.

## 5.2 Possible Approaches and Directions

**Possible research directions/approaches to tackle the challenge I:** In this paper, we wrote Python scripts to automatically identify pre-determined keywords and phrases, and used the WordNet database to look for synonyms, but this approach cannot detect the features that use synonyms not in the database. One approach to address this is to employ natural language processing (NLP) algorithms that can analyse contractual documentation and identify any non-compliant sections. These algorithms can be trained using annotated data to recognise specific language patterns to identify features and functionalities.

In [41] the authors used three Natural Language Processing tools (NLTK, Stanford CoreNLP, and spaCy) to identify, monitor and validate personally identifiable information in online contracts. In [42] the authors proposed an approach for extracting and classifying opt-out choices in websites' privacy policies, and the authors in [43] proposed an automated framework for privacy policy analysis based on a hierarchy of neural-network classifiers that accounts for both high-level aspects and fine-grained details of privacy practices. However, these are not designed for the analysis of the T & C and privacy agreement of mobile applications, and to the best of our knowledge, to date, no work has addressed the children data related regulations and safeguards.

**Possible research directions/approaches to tackle the challenge II:** Several papers propose static analysis approaches for verifying the security of Android apps (e.g., [49,50]), generating test cases (e.g., [51]), code efficiency check, private data leaks detection (e.g., [52, 53]), and detecting app clones (e.g., [54, 55]). The most common methods used for static analysis include (i) Abstract interpretation, (ii) taint analysis, and (iii) symbolic execution [44]. In the case of abstract interpretation, the codes are formally simplified (abstracted) and formal verification is used to prove the security, safety and functional properties of an app. The challenge with this approach is that formal abstraction is usually done manually, making full automation unlikely. In taint analysis, objects of interest (variable) are tainted and their values are tracked through data flow analysis. If a tainted object flows to an undesirable state then an alert is raised. To detect injection-type security attacks, variables that can be modified by users (e.g., inputs) are tainted and tracked. Symbolic execution generates possible program inputs and detects execution paths of interest. For example, it can be used to generate a sequence of GUI manipulations to identify the functionality of interest of an app (e.g., data transmission to detect information leakage) [56].

Our problem differs from the related works as we want to identify certain functions related to GDPR on children's data protection and safeguards. For example, we try to identify if a piece of code contains a function (or a set of functions) to collect geolocation data or a function that implements parental control or age verification. While a symbolic execution method similar to [56] could be applied, the meaning of the variables and functions is required to be defined. Combining symbolic execution, natural language processing, and/or abstract interpretation can be a potential approach.

Dynamic analysis can complement static analysis as the latter one may not cover all segments of the codes or execution traces in the case of large applications. A potential direction is to apply an automated testing approach similar to Appium [47], which tests the possible behaviour traces of an app. Screenshots of the GUI features during the operation of the application can be taken automatically. Then natural language processing, computer vision, and/or image recognition approaches can be used to identify and extract the relevant strings on the GUI of the app for assessing children's online safety. An advantage of this approach is that we only need to test the app as if a regular user would use it, instead of testing all possible inputs to detect malicious attacks, which avoids the behaviour traces explosion problem. However, this approach may require the analysis of a large number of pictures (screenshots) and an accurate classification of the features based on their semantics (meaning). For example, an extracted string "set display time" is likely to be part of parental control. To confirm this, static code analysis can be applied to the code segment containing this string (if not obfuscated).

Therefore, for identifying children safety-related features combining static and dynamic analysis seem to be a promising approach.

**Possible research directions/approaches to tackle the challenge III:** The app features and functions can be defined using different syntaxes such as a less formal XML-based language, or formal process algebra languages such as the applied π-calculus [57] and CSP [58]. However, these languages were not designed for modelling mobile app functions and features, therefore, they lack syntax for modelling parental control, age verification, and data storage/deletion. Other formal languages we can consider include the architecture language used in DataProve [48]. This language formally specifies the actions that a system architecture supports, such as receive, data collection, storage, deletion, and forward. However, it does not support parental control and age verification. Besides the syntax, to enable automated verification, the semantics (meaning) of the app features/functions should also be formalised.

**Possible research directions/approaches to tackle the challenge IV:** For specifying GDPR rules and safeguards policy languages can be used. There are numerous policy languages in the literature, such as the Preference Exchange Language (APPEL) [59] that enables web users to specify their privacy preferences that can be then matched against the practices set by the online services. The PrimeLife Privacy Policy Language (PPL) [60] enables the specification of access and usage control rules for the data subjects and the data controller. Its extended version, A-PPL [61], is an accountability policy language specifically designed for modelling data accountability, including data retention, log and notification. In [62], the Unified Modeling Language (UML) was extended to specify and represent different activities on data that can be checked for privacy compliance. The language called Privacy Enhanced Secure Tropos (PESTOS) was proposed in [63], which aids developers in catching GDPR privacy requirements at an early stage during their system design. Finally, the policy language used in DataProve [48] can be used to specify the end-to-end data protection policy of a service. These approaches lack syntax and semantics to model regulations and safeguards for children and can be extended for this purpose.

**Possible research directions/approaches to tackle the challenge V:** In general, in the area of automated verification, we distinguish between model-checking and theorem-proving approaches. Model-checking approaches [64] are usually based on verifying a finite state model of a system against a property and are automated. Theorem provers [65], on the other hand, are usually based on logic, and inference rules to assist mathematical proofs of safety and security properties of systems. To verify the compliance between the formally specified app features and GDPR rules and safeguard, we need to transform them into a common form. A promising approach applied by DataProve [48] is to transform two different specifications into first-order logic, which enables automated reasoning.

# 6 Discussion

This study focused entirely on children's data protection (as part of the EU GDPR) and also considers the EU and UK safeguarding guidelines in the mobile application context. Based on our analysis results, we found that despite the regulations and safeguards there are still many non-compliant applications, which can potentially be risky for children. Actually, the authors in 2018 [15] showed that a large number of applications in the US App Store potentially violated COPPA, and based on our results we can see that the situation after five years does not improve much (at least in the case of EU and UK regulations/safeguards). With the prospect of new and potentially stricter regulations on online safety, we hope to encourage further research and help researchers, developers better understand the issues with the safety of children using mobile apps.

For this analysis, we proposed a systematic method and procedure of five steps shown in Figure 2. In addition, based on our results, we suggest that there is a need for automated compliance check methods, and discussed the challenges and potential research directions in this area related to machine learning techniques, dynamic and static analysis, and formal specification and verification. We have demonstrated that automating aspects of the verification process could be possible while highlighting some practical difficulties due to the complex nature of the apps and the wide range of children-specific legislation and guidance that they must comply with. Nevertheless, we proposed an initial concept for automated compliance verification based on formal specifications. The automated non-compliancy detection could be the motivation and basis for designing educational and warning approaches designed specifically for children of certain age groups.

On the other hand, in this paper, we only focused on analysing Android applications for two reasons: (i) this work is part of a university-funded project called "*ChildDataVerif: Verifying the compliance of Android apps against regulations on children's data protection*", which focuses on Android applications, and (ii) the literature on Android app analysis and reverse engineering is more extensive than other platforms such as iOS apps. While there were some attempts to analyse iOS apps (e.g., [66,67]), doing it often requires jailbreaking, and there are still many other challenges and open questions [68].

In addition, in this study, we only address the seven features and functionalities given in Section 3.3.1. As a future direction, the study can be extended to examine additional features such as consent collection, data minimisation (which are covered in the GDPR), and online advertisements targetting children (as part of the Principle 1 in [8]), implementing privacy control features (Principle 5 in [8]), and provide education and awareness support (Principle 6 in [8]).

## 7  Conclusion and Future work

In this paper, we studied how Android applications comply with data protection regulations and online safeguards with regard to children's safety. We analysed a representative set of 91 Android applications against relevant data protection regulations and children's online safety guides. The analysis results show that the number of non-compliant apps is still significant, including a small set of apps designed for children. In preparing for the new regulations and bills (e.g., the Online Safety Bill in the UK), effective and accurate automated compliance verification approaches would be important to help app developers, designers and auditors to spot any issue at an early stage.

Addressing this, we proposed a possible approach for automated verification of mobile apps and identified the challenges and open questions. Possible future research related to these open questions includes (i) automated static and dynamic analysis methods designed for identifying features and functionalities related to children's safety and data protection, (ii) effective natural language processing methods can be designed to be used for the static and dynamic analysis, (iii) novel formal language to specify app features and children related to regulations and safeguards, and (iv) an efficient and accurate verification engine.

## Acknowledgements

This work and all authors have been supported by the project *ChildDataVerif: Verifying the compliance of Android apps against regulations on children's data protection*, funded under the Research Investment Fund, Edge Hill University, UK.

## References

[1] Connor Ibbetson. How many children have their own tech? https://yougov.co.uk/topics/education/articles-reports/2020/03/13/what-age-do-kids-get-phones-tablet-laptops-, 2020. Accessed: 2023-05-02.

[2] Jenny S. Radesky, Heidi M. Weeks, Rosa Ball, Alexandria Schaller, Samantha Yeo, Joke Durnez, Matthew Tamayo-Rios, Mollie Epstein, Heather L. Kirkorian, Sarah M. Coyne, and Rachel F. Barr. Young children's use of smartphones and tablets. *Pediatrics*, 146(1), July 2020.

[3] D. Smahel, H. Machackova, G. Mascheroni, L. Dedkova, E. Staksrud, K. Oĺafsson, S. Livingstone, and U. Hasebrink. Eu kids online 2020: Survey results from 19 countries. https://www.eukidsonline.ch/files/Eu-kids-online-2020-international-report.pdf, 2020. Accessed: 2023-05-02.

[4] Zoe Kleinman. Youtube accused of collecting uk children's data - bbc news. https://www.bbc.co.uk/news/technology-64786968, 3 2023. Accessed: 2023-01-24.

[5] Andrew Walker. Molly russell - prevention of future deaths report - 2022-0315. Technical report, North London Coroner's Service, 10 2022.

[6] NSPCC. Molly russell | nspcc. https://www.nspcc.org.uk/about-us/news-opinion/2022/response-molly-russell/, 9 2022. Accessed: 2023-01-29.

[7] Families Online. Warning to parents about zepeto after 'paedophile uses app to groom child'. https://www.familiesonline.co.uk/news/

urgent-warning-to-parents-about-zepeto-after-paedophile-uses-gaming-app-to-groom-child, 2 2022. Accessed: 2023-01-24.

[8] Social Media Working Group UKCCIS. Child safety online: A practical guide for providers of social media and interactive services. Report, 2015.

[9] Claudio Pollack. Child safety online: A practical guide for providers of social media and interactive services. https://www.gov.uk/government/publications/child-safety-online-a-practical-guide-for-providers-of-social-media-and-interactive-service 2016. Accessed: 2022-10-24.

[10] The European Parliament. Regulation (eu) 2016/679 of the european parliament and of the council of 27 april 2016 on the protection of natural persons with regard to the processing of personal data and on the free movement of such data, and repealing directive 95/46/ec (general data protection regulation). *Official Journal of the European Union*, 4 2016.

[11] ICO. Age appropriate design: a code of practice for online services. https://ico.org.uk/media/for-organisations/guide-to-data-protection/ico-codes-of-practice/age-appropriate-design-a-code-of-practice-for-online-services-2-1.pdf, 9 2020. Accessed: 2022-10-11.

[12] ICT Coalition. Home | ict coalition. http://www.ictcoalition.eu, 2022. Accessed: 2022-10-25.

[13] Federal Trade Commission. Children's online privacy protection rule: Not just for kids' sites | federal trade commission. https://www.ftc.gov/business-guidance/resources/childrens-online-privacy-protection-rule-not-just-kids-sites, 4 2023. Accessed: 2023-04-18.

[14] Sebastian Zimmeck, Ziqi Wang, Lieyong Zou, Roger Iyengar, Bin Liu, Florian Schaub, Shomir Wilson, Norman Sadeh, Steven Bellovin, and Joel Reidenberg. Automated analysis of privacy requirements for mobile apps. In *2016 AAAI Fall Symposium Series*, 2016.

[15] Irwin Reyes, Primal Wijesekera, Joel Reardon, Amit Elazari Bar On, Abbas Razaghpanah, Narseo Vallina-Rodriguez, and Serge Egelman. "won't somebody think of the children?" examining COPPA compliance at scale. *Proc. Priv. Enhancing Technol.*, 2018(3):63–83, 2018.

[16] Fangwei Zhao, Serge Egelman, Heidi M. Weeks, Niko Kaciroti, Alison L. Miller, and Jenny S. Radesky. Data Collection Practices of Mobile Applications Played by Preschool-Aged Children. *JAMA Pediatrics*, 174(12):e203345–e203345, 12 2020.

[17] Konrad Kollnig, Anastasia Shuba, Reuben Binns, Max Van Kleek, and Nigel Shadbolt. Are iphones really better for privacy? a comparative study of ios and android apps. *Proceedings on Privacy Enhancing Technologies*, 2022:6–24, 04 2022.

[18] Abdulaziz Aborujilah, Abdulaleem Z. Al-Othmani, Zalizah Awang Long, Nur Syahela Hussien, and Dahlan Abdul Ghani. Conceptual model for automating gdpr compliance verification using natural language approach. pages 1–6. IEEE, 12 2022.

[19] Suleiman Y. Yerima, Mohammed K. Alzaylaee, and Sakir Sezer. Machine learning-based dynamic analysis of android apps with improved code coverage. *EURASIP Journal on Information Security*, 2019:4, 12 2019.

[20] Hani Alshahrani, Harrison Mansourt, Seaver Thorn, Ali Alshehri, Abdulrahman Alzahrani, and Huirong Fu. Ddefender: Android application threat detection using static and dynamic analysis. pages 1–6. IEEE, 1 2018.

[21] Zia Muhammad, Faisal Amjad, Zafar Iqbal, Abdul Rehman Javed, and Thippa Reddy Gadekallu. Circumventing google play vetting policies: a stealthy cyberattack that uses incremental updates to breach privacy. *Journal of Ambient Intelligence and Humanized Computing*, 1 2023.

[22] Arup Kumar Ghosh, Karla Badillo-Urquiola, Shion Guha, Joseph J. LaViola Jr, and Pamela J. Wisniewski. Safety vs. surveillance. pages 1–14. ACM, 4 2018.

[23] Bing Hu, Bin Liu, Neil Zhenqiang Gong, Deguang Kong, and Hongxia Jin. Protecting your children from inappropriate content in mobile apps. pages 1111–1120. ACM, 10 2015.

[24] Rocco Zaccagnino, Carmine Capo, Alfonso Guarino, Nicola Lettieri, and Delfina Malandrino. Techno-regulation and intelligent safeguards. *Multimedia Tools and Applications*, 80:15803–15824, 4 2021.

[25] Bibi van den Berg and Robert Leenes. Abort, retry, fail: scoping techno-regulation and other techno-effects. *Human law and computer law: comparative perspectives*, pages 67–87, 2013.

[26] Google Search. Google. https://www.google.com, 2022. Accessed: 2022-10-24.

[27] Catherine Brown. The most popular apps for teenagers: What's hot, what's not. https://yourteenmag.com/technology/the-most-popular-apps-for-teenagers, 11 2017. Accessed: 2023-03-06.

[28] fyiplayitsafe. Just released! most popular apps for teenagers in 2021. https://fyiplayitsafe.com/just-released-most-popular-apps-for-teenagers-in-2021/, 2022. Accessed: 2023-10-24.

[29] besociallysmart. Be socially smart. https://besociallysmart.com/top-apps-teens/, 2022. Accessed: 2023-01-24.

[30] SerpApi. Google play store api - serpapi. https://serpapi.com/google-play-api, 2023. Accessed: 2023-01-24.

[31] Google Play Help. Apps & games content ratings on google play - google play help. https://support.google.com/googleplay/answer/6209544, 2023. Accessed: 2023-04-02.

[32] Legislation.gov.uk. Data protection act 2018. https://www.legislation.gov.uk/ukpga/2018/12/contents/enacted, 5 2018. Accessed: 2023-04-18.

[33] Google Play. Teacher approved | google play console. https://play.google.com/console/about/programs/teacherapproved/, 2023. Accessed: 2023-01-26.

[34] Google Developers. About android app bundles | android developers. https://developer.android.com/guide/app-bundle, 2023. Accessed: 2023-03-30.

[35] APK Downloader. Apk downloader [latest] download directly | january 2023 | (evozi official). https://apps.evozi.com/apk-downloader/, 2020. Accessed: 2023-01-24.

[36] APKPure. Download apk fast, free and safe on android. https://m.apkpure.com/, 2023. Accessed: 2023-01-24.

[37] APK-DL. Apk-dl : Download android apk files. https://apk-dl.com/, 2023. Accessed: 2023-01-24.

[38] APKMirror. Apkmirror - free apk downloads - free and safe android apk downloads. https://www.apkmirror.com/, 2023. Accessed: 2023-01-24.

[39] APKTool. Apktool - a tool for reverse engineering 3rd party, closed, binary android apps. https://ibotpeaches.github.io/Apktool/, 2023. Accessed: 2023-01-24.

[40] Princeton University. Wordnet. https://wordnet.princeton.edu/, 2023. Accessed: 2023-03-20.

[41] Paulo Silva, Carolina Gonçalves, Carolina Godinho, Nuno Antunes, and Marilia Curado. Using natural language processing to detect privacy violations in online contracts. In *Proceedings of the 35th Annual ACM Symposium on Applied Computing*, SAC '20, page 1305–1307, New York, NY, USA, 2020. Association for Computing Machinery.

[42] Vinayshekhar Bannihatti Kumar, Roger Iyengar, Namita Nisal, Yuanyuan Feng, Hana Habib, Peter Story, Sushain Cherivirala, Margaret Hagan, Lorrie Cranor, Shomir Wilson, Florian Schaub, and Norman Sadeh. Finding a choice in a haystack: Automatic extraction of opt-out statements from privacy policy text. WWW '20, page 1943–1954, New York, NY, USA, 2020. Association for Computing Machinery.

[43] Hamza Harkous, Kassem Fawaz, Rémi Lebret, Florian Schaub, Kang G. Shin, and Karl Aberer. Polisis: Automated analysis and presentation of privacy policies using deep learning. In *Proceedings of the 27th USENIX Conference on Security Symposium*, SEC'18, page 531–548, USA, 2018. USENIX Association.

[44] Janaka Senanayake, Harsha Kalutarage, Mhd Omar Al-Kadri, Andrei Petrovski, and Luca Piras. Android source code vulnerability detection: A systematic literature review. *ACM Comput. Surv.*, 55(9), jan 2023.

[45] Li Li, Tegawendé F. Bissyandé, Mike Papadakis, Siegfried Rasthofer, Alexandre Bartel, Damien Octeau, Jacques Klein, and Le Traon. Static analysis of android apps: A systematic literature review. *Information and Software Technology*, 88:67–95, 2017.

[46] V Benjamin Livshits and Monica S Lam. Finding security vulnerabilities in java applications with static analysis. In *USENIX security symposium*, volume 14, pages 18–18, 2005.

[47] Junmei Wang and Jihong Wu. Research on mobile application automation testing technology based on appium. In *2019 International Conference on Virtual Reality and Intelligent Systems (ICVRIS)*, pages 247–250, 2019.

[48] Vinh Thong Ta and Max Hashem Eiza. Dataprove: Fully automated conformance verification between data protection policies and system architectures. *Proceedings on Privacy Enhancing Technologies (PoPETs)*, 2022(1):565–585, January 2022.

[49] Yajin Zhou and Xuxian Jiang. Detecting passive content leaks and pollution in android applications. In *Network and Distributed System Security Symposium*, 2013.

[50] Damien Octeau, Daniel Luchaup, Matthew L. Dering, Somesh Jha, and Patrick Mcdaniel. Composite constant propagation: Application to android inter-component communication analysis. *2015 IEEE/ACM 37th IEEE International Conference on Software Engineering*, 1:77–88, 2015.

[51] Nariman Mirzaei, Hamid Bagheri, Riyadh Mahmood, and Sam Malek. Sig-droid: Automated system input generation for android applications. pages 461–471, 11 2015.

[52] Steven Arzt, Siegfried Rasthofer, Christian Fritz, Eric Bodden, Alexandre Bartel, Jacques Klein, Yves Le Traon, Damien Octeau, and Patrick McDaniel. Flowdroid: Precise context, flow, field, object-sensitive and lifecycle-aware taint analysis for android apps. *Acm Sigplan Notices*, 49(6):259–269, 2014.

[53] Vitalii Avdiienko, Konstantin Kuznetsov, Alessandra Gorla, Andreas Zeller, Steven Arzt, Siegfried Rasthofer, and Eric Bodden. Mining apps for abnormal usage of sensitive data. In *Proceedings of the 37th International Conference on Software Engineering - Volume 1*, ICSE '15, page 426–436. IEEE Press, 2015.

[54] Jonathan Crussell, Clint Gibler, and Hao Chen. Attack of the clones: Detecting cloned applications on android markets. In Sara Foresti, Moti Yung, and Fabio Martinelli, editors, *Computer Security – ESORICS 2012*, pages 37–54, Berlin, Heidelberg, 2012. Springer Berlin Heidelberg.

[55] Jonathan Crussell, Clint Gibler, and Hao Chen. Andarwin: Scalable detection of semantically similar android applications. In Jason Crampton, Sushil Jajodia, and Keith Mayes, editors, *Computer Security – ESORICS 2013*, pages 182–199, Berlin, Heidelberg, 2013. Springer Berlin Heidelberg.

[56] Zhemin Yang, Min Yang, Yuan Zhang, Guofei Gu, Peng Ning, and X. Sean Wang. Appintent: Analyzing sensitive data transmission in android for privacy leakage detection. In *Proceedings of the 2013 ACM SIGSAC Conference on Computer Communications Security*, CCS '13, page 1043–1054, New York, NY, USA, 2013. Association for Computing Machinery.

[57] Martín Abadi, Bruno Blanchet, and Cédric Fournet. The applied pi calculus: Mobile values, new names, and secure communication. *J. ACM*, 65(1), oct 2017.

[58] Stephen D. Brookes and A.W. Roscoe. *CSP: A Practical Process Algebra*, page 187–222. Association for Computing Machinery, New York, NY, USA, 1 edition, 2021.

[59] The Platform for Privacy Preferences (P3P). APPEL 1.0, 2012. http://www.w3.org/TR/2002/WD-P3P-preferences-20020415/.

[60] S Trabelsi, Akram Njeh, Laurent Bussard, and Gregory Neven. Ppl engine: A symmetric architecture for privacy policy handling. *W3C Workshop on Privacy and data usage control*, pages 1–5, 04 2010.

[61] Monir Azraoui, Kaoutar Elkhiyaoui, Melek Önen, Karin Bernsmed, Anderson Santana De Oliveira, and Jakub Sendor. A-ppl: An accountability policy language. In Joaquin Garcia-Alfaro, Jordi Herrera-Joancomartí, Emil Lupu, Joachim Posegga, Alessandro Aldini, Fabio Martinelli, and Neeraj Suri, editors, *Data Privacy Management, Autonomous Spontaneous Security, and Security Assurance*, pages 319–326, Cham, 2015. Springer.

[62] M. Alshammari and A. Simpson. A model-based approach to support privacy compliance. *Information and Computer Security,*, 26(4):437–453, 2018.

[63] I. Çelebi. Privacy enhanced secure tropos : A privacy modeling language for gdpr compliance. Master Thesis, 2018.

[64] Ranjit Jhala and Rupak Majumdar. Software model checking. *ACM Comput. Surv.*, 41(4), oct 2009.

[65] M. Saqib Nawaz, Moin Malik, Yi Li, Meng Sun, and Muhammad Ikram Ullah Lali. A survey on theorem provers in formal methods. *CoRR*, abs/1912.03028, 2019.

[66] Manuel Egele, Christopher Kruegel, Engin Kirda, and Giovanni Vigna. Pios: Detecting privacy leaks in ios applications. In *NDSS*, pages 177–183, 2011.

[67] Daniil Tiganov, Jeff Cho, Karim Ali, and Julian Dolby. Swan: A static analysis framework for swift. In *Proceedings of the 28th ACM Joint Meeting on European Software Engineering Conference and Symposium on the Foundations of Software Engineering*, ESEC/FSE 2020, page 1640–1644, New York, NY, USA, 2020. Association for Computing Machinery.

[68] Martin Szydlowski, Manuel Egele, Christopher Kruegel, and Giovanni Vigna. Challenges for dynamic analysis of ios applications. In Jan Camenisch and Dogan Kesdogan, editors, *Open Problems in Network Security*, pages 65–77, Berlin, Heidelberg, 2012. Springer Berlin Heidelberg.